\documentclass{phyeauth}
\usepackage{graphicx}
\usepackage{amsmath}
\usepackage{amssymb}

\newcommand{\llangle}{\langle\!\langle}
\newcommand{\rrangle}{\rangle\!\rangle}
\newcommand{\rlangle}{\rrangle\!\llangle}

\begin{document}

\begin{frontmatter}
\title{Current noise spectrum of a quantum shuttle}

\author[mic]{Christian Flindt\thanksref{thank1}},
\author[ku,prague]{Tom\'{a}\v{s} Novotn\'{y}} and
\author[mic]{Antti-Pekka Jauho}

\address[mic]{MIC -- Department of Micro and Nanotechnology, Technical University of Denmark,
  DTU - Building 345east, DK-2800~Kongens~Lyngby, Denmark}
\address[ku]{Nano-Science Center, University of Copenhagen - Universitetsparken 5, DK-2100~Copenhagen \O, Denmark}
\address[prague]{Department of Electronic Structures, Faculty of Mathematics and Physics,
  Charles University - Ke Karlovu 5, 12116~Prague, Czech Republic}

\thanks[thank1]{Corresponding author. E-mail: cf@mic.dtu.dk}

\begin{abstract}
We present a method for calculating the full current noise
spectrum $S(\omega)$ for the class of nano-electromechanical
systems (NEMS) that can be described by a Markovian generalized
master equation. As a specific example we apply the method to a
quantum shuttle. The noise spectrum of the shuttle has peaks at
integer multiples of the mechanical frequency, which is slightly
renormalized. The renormalization explains a previously observed
small deviation of the shuttle current compared to the expected
value given by the product of the natural mechanical frequency and
the electron charge.  For a certain parameter range the quantum
shuttle exhibits a coexistence regime, where the charges are
transported by two different mechanisms: shuttling and sequential
tunneling. In our previous studies we showed that characteristic
features in the zero-frequency noise could be quantitatively
understood as a slow switching process between the two current
channels, and the present study shows that this interpretation
holds also qualitatively at finite frequency.
\end{abstract}

\end{frontmatter}

\section{Introduction}
A decade of advances in microfabrication technology has pushed the
typical length scales of electromechanical systems to the limit,
where quantum mechanical effects of the mechanical motion must be
taken into account \cite{blencowe_2004}. Such
nano-electromechanical systems (NEMS) exhibit a strong interplay
between mechanical and electronic (or magnetic) degrees of
freedom, and their electronic transport properties reflect this
interplay in an intricate manner.

A modern trend in transport studies of mesoscopic systems has been
to not only consider the current-voltage characteristics of a
given device, but also to examine the noise properties, or even
the higher cumulants (\emph{i.e.} the full counting statistics
(FCS)) of the current distribution
\cite{blanter_2000,nazarov_2003}. The current noise, either its
zero-frequency component or the whole frequency spectrum, provides
more information than just the mean current and can be used to
discern among different possible mechanisms resulting in the same
mean current. While the noise spectra in generic mesoscopic
systems have been studied well over a decade, it is only very
recently that the study of noise spectra of NEMS has been
initiated \cite{isacsson_2004,armour_2004,koch_preprint}.

The aforementioned three studies deal with the noise spectra of a
classical shuttle, a classical nanomechanical resonator coupled to
a single electron transistor (SET), and the C$_{60}$-SET in a
strong electromechanical coupling regime, respectively. The first
two studies \cite{isacsson_2004,armour_2004} found peaks in the
current noise spectra at the first two multiples of the mechanical
frequency. For low bias voltages and strong electromechnical
coupling the third study \cite{koch_preprint} found a power-law
frequency dependence of the noise spectrum attributed to
scale-free avalanche charge transfer processes. In all three cases
the noise spectra revealed interesting details about the systems.
From the technical point of view, two of the studies
\cite{isacsson_2004,koch_preprint} used Monte Carlo simulations,
whereas \cite{armour_2004} used a model-specific numerical
evaluation of the MacDonald formula (see below).

In this work, we present a study of the full frequency spectrum of
the current noise of a quantum shuttle
\cite{gorelik_1998,novotny_2003,fedorets_2004}. We extend the
general formalism developed for the zero-frequency noise
\cite{novotny_2004,flindtPRB_2004} and the FCS
\cite{flindtEPL_2004} calculations for NEMS described by a
Markovian generalized master equation. The presented formalism
applies not only to the shuttle studied here but could equally
well be used for all three systems from the previous studies
\cite{isacsson_2004,armour_2004,koch_preprint} for the
determination of the noise spectra.

We apply the developed theory to compute numerically the noise
spectrum of the shuttle in the deep quantum regime\footnote{In
this regime $\lambda \simeq x_0$, where $\lambda$ is the tunneling
length (see also Eq.(\ref{eq_ngme})) and
$x_0=\sqrt{\hbar/m\omega_0}$ sets the length scale for the quantum
mechanical zero-point motion.} as function of the damping
coefficient. The spectrum has peaks at integer multiples of the
slightly {\em renormalized} mechanical frequency. The
renormalization of the bare oscillator frequency as read off from
the current spectrum explains a small but observable deviation
from the expected value of the current in the shuttling regime
$I_{\rm shut}=e\omega_0/2\pi$ \cite{novotny_2003}. It turns out
that it is the renormalized oscillator frequency
$\tilde{\omega}_0$ which should enter this relation. Finally, we
focus on the low-frequency part of the spectrum when approaching
the semi-classical regime for intermediate values of the damping,
\emph{i.e.} in the coexistence regime, where both shuttling and
tunneling are effective. We use the frequency dependence of the
spectrum for $\omega\ll \omega_0$ to identify additional
qualitative evidence for the bistable behavior of the shuttle in
this regime described by a simple analytical theory of a slow
switching between two current channels (compare with Refs.\
\cite{flindtEPL_2004,andrea_thesis}).

\section{Model}
\begin{figure}
  \centering
  \includegraphics[width=0.47\textwidth]{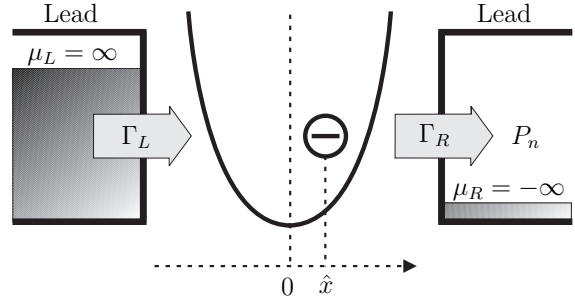}
  \caption{The quantum shuttle consists of a
  nanosized grain moving in a harmonic potential between two leads.
  A high bias between the leads drives electrons through the grain.}
  \label{fig_setup}
\end{figure}

We consider the model of a quantum shuttle described in
\cite{gorelik_1998,novotny_2003,fedorets_2004,novotny_2004}. The
shuttle consists of a mechanically oscillating nanoscale grain
situated between two leads (see Fig. \ref{fig_setup}). In the
strong Coulomb blockade regime the grain can be treated as having
a single electronic level only. A high bias between the leads
drives electrons through the grain and exerts an electrostatic
force on the grain, when charged. The grain is assumed to move in
a harmonic potential, and the oscillations of the grain are
treated fully quantum mechanically. Damping of the oscillations is
described by interactions with a surrounding heat bath.

From the Hamiltonian of the model one can derive a generalized
master equation (GME) resolved with respect to the number of
electrons $n$ that have been collected in the right lead during
the time span $0$ to $t$. The $n$-resolved GME describes the time
evolution of the $n$-resolved system density matrix
$\hat{\rho}^{(n)}(t)$, where the `system' consists of the
electronic level of the grain and the quantized oscillations. In
the following we only need the $n$-resolved GME for the part of
$\hat{\rho}^{(n)}(t)$ that is diagonal in the electronic
components, which reads \cite{novotny_2004}

\begin{equation}
\begin{split}
\dot{\hat{\rho}}_{00}^{(n)}(t)=&
\frac{1}{i\hbar}[\hat{H}_{\mathrm{osc}},\hat{\rho}_{00}^{(n)}(t)]
+\mathcal{L}_{\mathrm{damp}}\hat{\rho}_{00}^{(n)}(t)\\
&-\frac{\Gamma_L}{2}\{e^{-\frac{2\hat{x}}{\lambda}},\hat{\rho}_{00}^{(n)}(t)\}
+\Gamma_Re^{\frac{\hat{x}}{\lambda}}\hat{\rho}_{11}^{(n-1)}(t)e^{\frac{\hat{x}}{\lambda}},\\
\dot{\hat{\rho}}_{11}^{(n)}(t)=&
\frac{1}{i\hbar}[\hat{H}_{\mathrm{osc}}-eE\hat{x},\hat{\rho}_{11}^{(n)}(t)]
+\mathcal{L}_{\mathrm{damp}}\hat{\rho}_{11}^{(n)}(t)\\
&-\frac{\Gamma_R}{2}\{e^{\frac{2\hat{x}}{\lambda}},\hat{\rho}_{11}^{(n)}(t)\}
+\Gamma_Le^{-\frac{\hat{x}}{\lambda}}\hat{\rho}_{00}^{(n)}(t)e^{-\frac{\hat{x}}{\lambda}},
\end{split}
\label{eq_ngme}
\end{equation}
with $n=0,1,2,\ldots$ and $\hat{\rho}_{11}^{(-1)}(t)\equiv 0$. The
commutators describe the coherent evolution of the charged
($\hat{\rho}_{11}^{(n)}\equiv\langle 1|\hat{\rho}^{(n)}|1\rangle$)
or empty ($\hat{\rho}_{00}^{(n)}\equiv\langle
0|\hat{\rho}^{(n)}|0\rangle$) shuttle with mass $m$ and natural
frequency $\omega_0$. The electric field\footnote{In order to
obtain a Markovian GME it is necessary to assume that the bias
applied between the leads is the highest energy scale of the
system \cite{novotny_2003}. Nevertheless, we consider the electric
field between the leads as a free parameter of the model.} between
the leads is denoted $E$. The terms proportional to
$\Gamma_{L(R)}$ describe transfer processes from the left (to the
right) lead with hopping amplitudes that depend exponentially on
the ratio between position $\hat{x}$ and the electron tunneling
length $\lambda$. The mechanical damping of the oscillator is
described by the damping kernel (here at zero temperature)
$\mathcal{L}_{\rm{damp}}\hat{\rho}^{(n)}_{jj} =
-\frac{i\gamma}{2\hbar}[\hat{x},\{\hat{p},\hat{\rho}^{(n)}_{jj}\}]-\frac{\gamma
m\omega}{2\hbar}[\hat{x},[\hat{x},\hat{\rho}^{(n)}_{jj}]], j=0,1$
\cite{novotny_2003,novotny_2004}.

The $n$-resolved GME can be recast into the compact form
\cite{flindtPRB_2004}
\begin{equation}
\dot{\hat{\rho}}^{(n)}=(\mathcal{L}-\mathcal{I}_R)\hat{\rho}^{(n)}+\mathcal{I}_R\hat{\rho}^{(n-1)},\hat{\rho}^{(-1)}\equiv
0, \label{eq_ngme_compact}
\end{equation}
where we have introduced the Liouvillean $\mathcal{L}$, describing
the evolution of the system density matrix
$\hat{\rho}(t)=\sum_n\hat{\rho}^{(n)}(t)$, \emph{i.e.}
$\dot{\hat{\rho}}(t)=\mathcal{L}\hat{\rho}(t)$, and the
superoperator for the tunnel current through the right junction
(taking $e=1$), defined by its action on the density operator
\begin{equation}
\mathcal{I}_R\hat{\rho} =
\Gamma_Re^{\frac{\hat{x}}{\lambda}}|0\rangle\!\langle
1|\hat{\rho}|1\rangle\!\langle 0| e^{\frac{\hat{x}}{\lambda}}.
\end{equation}

Assuming that the system tends exponentially  to a stationary
state $\hat{\rho}^{\mathrm{stat}}$ the Liouvillean has a single
eigenvalue equal to zero with $\hat{\rho}^{\mathrm{stat}}$ being
the (unique and normalized) right eigenvector which we denote by
$|0\rrangle$ \cite{flindtPRB_2004}. The corresponding left
eigenvector is the identity operator $\hat{1}$ which we denote by
$\llangle\tilde{0}|$, and from the definition of the inner
product\footnote{We define the inner product of two supervectors
as $\llangle a|b\rrangle=\mathrm{Tr}(\hat{A}^{\dagger}\hat{B})$
with the identification $|o\rrangle\leftrightarrow \hat{O}$, where
$\hat{O}$ is a quantum mechanical operator and $|o\rrangle$ is the
corresponding supervector.} we have
$\llangle\tilde{0}|0\rrangle=\mathrm{Tr}(\hat{1}^{\dagger}\hat{\rho}^{\mathrm{stat}})=1$.
In terms of $\mathcal{I}_R$ the average tunnel current in the
stationary state can be expressed as
\begin{equation}
\langle \hat{I}_R\rangle
=\mathrm{Tr}(\mathcal{I}_R\hat{\rho}^{\mathrm{stat}}) = \llangle
\tilde{0}|\mathcal{I}_R|0\rrangle.
\end{equation}
We define the projectors $\mathcal{P}=|0\rlangle\tilde{0}|$ and
$\mathcal{Q}=1-\mathcal{P}$ obeying the relations
$\mathcal{PL}=\mathcal{LP}=0$ and $\mathcal{QLQ}=\mathcal{L}$. In
terms of the two projectors we can express the resolvent of the
Liouvillean $\mathcal{G}(-i\omega)=(-i\omega-\mathcal{L})^{-1}$ as
\begin{equation}
\mathcal{G}(-i\omega)=-\frac{1}{i\omega}\mathcal{P}
-\mathcal{Q}\frac{1}{i\omega+\mathcal{L}}\mathcal{Q}=-\frac{1}{i\omega}\mathcal{P}-\mathcal{R}(\omega),
\label{eq_resolvent}
\end{equation}
where we have introduced the frequency dependent superoperator
$\mathcal{R}(\omega)$, which is well-defined even for $\omega=0$,
since the inversion in that case is performed only in the subspace
where $\mathcal{L}$ is regular.

\section{Theory}

We consider the current autocorrelation function defined as
\begin{equation}
C_{II}(t', t'')= \frac{1}{2}\langle\{\Delta \hat{I}(t'),\Delta
\hat{I}(t'') \}\rangle,
\end{equation}
where $\Delta \hat{I}(t)=\hat{I}(t)-\langle \hat{I}(t)\rangle$. In
the stationary state $C_{II}(t', t'')$ can only be a function of
the time difference $t=t'-t''$, and we thus write
\begin{equation}
C_{II}(t)= \frac{1}{2}\langle\{\Delta \hat{I}(t),\Delta \hat{I}(0)
\}\rangle.
\end{equation}
The current noise spectrum is the Fourier transform of
$C_{II}(t)$, \emph{i.e.}
\begin{equation}
S_{II}(\omega)\equiv\int_{-\infty}^{\infty}dt C_{II}(t)e^{i\omega
t}.
\end{equation}
In order to calculate the current noise measurable in, say, the
right lead one must recognize that the current running in the lead
is a sum of two contributions, namely the tunnel current through
the right junction and a displacement current induced by electrons
tunneling between leads and grain. This is reflected in the
Ramo-Shockley theorem~\cite{blanter_2000}
\begin{equation}
\hat{I}=c_L\hat{I}_R+c_R\hat{I}_L.
\label{eq_ramo-shockly}
\end{equation}
Here $\hat{I}$ is the current operator for the current running in
the lead, whereas $\hat{I}_{L(R)}$ is the operator for the tunnel
current through the left (right) junction, and $c_{L(R)}$ is the
relative capacitance of the left (right) junction in the sense
$c_L+c_R=1$. Combining the Ramo-Shockley theorem with charge
conservation leads to an expression for the current noise measured
in the lead~\cite{aguado_2004}
\begin{equation}
S_{II}(\omega)=c_LS_{I_R I_R}(\omega)+c_R
S_{I_LI_L}(\omega)-c_Lc_R\omega^2S_{NN}(\omega),
\label{eq_fullnoise}
\end{equation}
where $\hat{N}=|1\rangle\!\langle 1|$ is the occupation number
operator of the electronic level of the grain. In the following we
neglect any dependence of the capacitances on the position of the
grain and consider the symmetric case $c_L=c_R=\frac{1}{2}$.

The two first terms of Eq. (\ref{eq_fullnoise}) can be evaluated
using the methods developed by MacDonald\cite{macdonald_1962}. The
starting point of the derivation is the property
$C_{II}(t)=C_{II}(-t)$, which immediately leads to
\begin{equation}
S_{I_RI_R}(\omega)=\int_0^{\infty}dt C_{I_RI_R}(t)(e^{i\omega t
}+e^{-i\omega t}).
\end{equation}
Let us consider the first term
\begin{equation}
S^{+}_{I_RI_R}(\omega)\equiv\int_0^{\infty}dt
C_{I_RI_R}(t)e^{i\omega t}, \label{eq_Splus}
\end{equation}
the second term, $S^{-}_{I_RI_R}(\omega)$, follows analogously.
Defining $\hat{Q}_R(t)$ as the operator of charge collected in the
right lead in the time span $0$ to $t$ we have
\begin{equation}
\Delta
\hat{Q}_R(t)=\hat{Q}_R(t)-\langle\hat{Q}_R(t)\rangle=\int_{0}^{t}dt'\Delta
\hat{I}_R(t'),
\end{equation}
and we can express the current autocorrelation function as
\begin{equation}
C_{I_RI_R}(t)=\frac{1}{2} \frac{d}{dt}\langle\{\Delta
\hat{Q}_R(t), \Delta\hat{I}_R(0) \}\rangle.
\end{equation}
Introducing the convergence factor $\varepsilon\rightarrow 0^+$
and performing the integration by parts in Eq.\ (\ref{eq_Splus})
we get
\begin{equation}
\begin{split}
S^{+}_{I_RI_R}(\omega)=\int_0^{\infty}&dt\langle\{\Delta
\hat{Q}_R(t), \Delta\hat{I}_R(0) \}\rangle\\
&\times\frac{\omega+i\varepsilon}{2i}e^{i(\omega+i\varepsilon)t}.
\end{split}
\end{equation}
Since $\langle\{\Delta \hat{Q}_R(t), \Delta\hat{I}_R(0)
\}\rangle=\langle\{\Delta \hat{Q}_R(t), \Delta\hat{I}_R(t)
\}\rangle=\frac{d}{dt}\langle\Delta\hat{Q}^2_R(t)\rangle$ in the
stationary state\footnote{The first equality follows from a simple
substitution $\tau\to t-\tau$ in the chain $\langle\{\Delta
\hat{Q}_R(t), \Delta\hat{I}_R(0)\}\rangle=\int_0^t
d\tau\langle\{\Delta \hat{I}_R(\tau), \Delta\hat{I}_R(0)
\}\rangle=\int_0^t d\tau\langle\{\Delta \hat{I}_R(t),
\Delta\hat{I}_R(\tau)\}\rangle=\langle\{\Delta \hat{Q}_R(t),
\Delta\hat{I}_R(t) \}\rangle$.} we can write
\begin{equation}
S^{+}_{I_RI_R}(\omega)=\int_0^{\infty}dt\frac{d}{dt}\langle\Delta
\hat{Q}^2_R(t)\rangle\frac{\omega+i\varepsilon}{2i}e^{i(\omega+i\varepsilon)t}.
\end{equation}
Similarly, we find
\begin{equation}
\begin{split}
S^{-}_{I_RI_R}(\omega)=-\int_0^{\infty}dt\frac{d}{dt}\langle\Delta
\hat{Q}^2_R(t)\rangle\frac{\omega-i\varepsilon}{2i}e^{-i(\omega-i\varepsilon)t},
\end{split}
\end{equation}
and consequently
\begin{equation}
S_{I_RI_R}(\omega)=\int_0^{\infty} dt\frac{d}{dt}\langle\Delta
\hat{Q}^2_R(t)\rangle (\omega\sin{\omega t}+\varepsilon\cos{\omega
t})e^{-\varepsilon t}.
\end{equation}
We now make use of the fact that
\begin{equation}
\langle\Delta \hat{Q}^2_R(t)\rangle=\langle
\hat{Q}^2_R(t)\rangle-\langle \hat{Q}_R(t)\rangle^2=\langle
n^2(t)\rangle-\langle n(t)\rangle^2,
\end{equation}
with $\langle n^{\alpha}(
t)\rangle\equiv\sum_{n=0}^{\infty}n^{\alpha}P_n( t), \alpha=1,2$,
where $P_n( t)$ is the probability of having collected
$n=0,1,2,\ldots$ electrons in the right lead during the time span
$0$ to $t$. This finally leads us to the commonly used form of the
MacDonald formula (cf.\
\cite{armour_2004,aguado_2004,macdonald_1962})
\begin{equation}
S_{I_RI_R}(\omega)= \omega \int_{0}^{\infty}dt\sin(\omega
t)\frac{d}{dt}\left[\langle n^2( t)\rangle-\langle
n(t)\rangle^2\right], \label{eq_macdonald}
\end{equation}
where the regularization
\begin{equation}
\begin{split}
\omega\sin(\omega t)&\to\frac{1}{2i}
\big((\omega+i\varepsilon)e^{i(\omega+i\varepsilon)t}
-(\omega-i\varepsilon)e^{-i(\omega-i\varepsilon)t}\big)\\
&=(\omega\sin{\omega t}+\varepsilon\cos{\omega t})e^{-\varepsilon
t},\varepsilon\rightarrow 0^{+}
\end{split}
\end{equation}
is implied. Only the proper treatment of the regularization
ensures correct results including the $\omega=0$ case where the
zero-frequency noise MacDonald formula is recovered (by using the
Laplace transform identity for $\varepsilon\to 0^+$)
\begin{equation}
\begin{split}
S_{I_RI_R}(0)&= \varepsilon\int_{0}^{\infty}dt\, e^{-\varepsilon
t}\frac{d}{dt}\left[\langle n^2(
t)\rangle-\langle n(t)\rangle^2\right]\\
&=\frac{d}{dt}\left[\langle n^2( t)\rangle-\langle
n(t)\rangle^2\right]\Big|_{t\rightarrow \infty}.
\end{split}
\end{equation}

In order to evaluate the current noise spectrum we now introduce
the quantity
\begin{equation}
\begin{split}
&\tilde{S}(\omega) = \omega\int_{0}^{\infty}dte^{i\omega
t}\left[\frac{d}{dt}\langle n^2(t)\rangle-2\langle n(t)\rangle
\frac{d}{dt}\langle n(t)\rangle\right] \label{eq_squantity}
\end{split}
\end{equation}
with either $S(\omega)={\rm Im}\tilde{S}(\omega)$ or
$S(\omega)=\big(\tilde{S}(\omega)+\tilde{S}(-\omega)\big)/2i$ and
evaluate it along the lines of \cite{flindtPRB_2004}. Since
$P_n(t)=\mathrm{Tr}(\hat{\rho}^{(n)}(t))$, Eq.
(\ref{eq_ngme_compact}) leads to (keeping in mind that
$\mathrm{Tr}(\mathcal{L}\bullet)=0$)
\begin{align}
&\dot{P}_n(t)=\mathrm{Tr}[\mathcal{I}_R(\hat{\rho}^{(n-1)}(t)-\hat{\rho}^{(n)}(t))],\label{eq_sum0}
\intertext{and as shown in \cite{flindtPRB_2004}}
&\frac{d}{dt}\langle n(t)\rangle=\mathrm{Tr}(\mathcal{I}_R\hat{\rho}(t))=\llangle\tilde{0}|\mathcal{I}_R|0\rrangle,\label{eq_sum1}\\
&\frac{d}{dt}\langle n^2(t)\rangle=
2\mathrm{Tr}\big[\mathcal{I}_R\sum_nn\hat{\rho}^{(n)}(t)\big]+\llangle\tilde{0}|\mathcal{I}_R|0\rrangle,\label{eq_sum2}
\end{align}
where we have used $\hat{\rho}(t)=\hat{\rho}^{\mathrm{stat}}$,
since we are considering the stationary limit. The sum entering
Eq. (\ref{eq_sum2}) is evaluated by introducing an operator-valued
generating function defined as
$\hat{F}(t,z)=\sum_{n=0}^{\infty}\hat{\rho}^{(n)}(t)z^n$, from
which we get
\begin{equation}
\frac{\partial}{\partial
z}\hat{F}(t,z)\Big|_{z=1}=\sum_nn\hat{\rho}^{(n)}(t).
\label{eq_diffF}
\end{equation}
From the definition of the Laplace transform,
$\tilde{\hat{F}}(s,z)=\int_0^{\infty}dt\hat{F}(t,z)e^{-st}$, we
see that the integration in Eq. (\ref{eq_squantity}) can be
considered as a Laplace transform evaluated at
$s=-i\omega+\varepsilon$ (remember the proper regularization; from
now on we skip explicitly mentioning the $\varepsilon$-factors).
In \cite{flindtPRB_2004} it was shown that
\begin{equation}
\begin{split}
\frac{\partial}{\partial
z}\tilde{\hat{F}}(s=-i\omega,z)\Big|_{z=1}=&\mathcal{G}(-i\omega)\mathcal{I}_R\mathcal{G}(-i\omega)\hat{\rho}(0)\\
&+\mathcal{G}(-i\omega)\sum_nn\hat{\rho}^{(n)}(0).
\end{split}
\label{eq_d/dzF}
\end{equation}
Again, we have $\hat{\rho}(0)=\hat{\rho}^{\mathrm{stat}}$, and
moreover we assume the factorized initial condition
\cite{aguado_2004,korotkov_2003}
$\hat{\rho}^{(n)}(0)=\delta_{0n}\hat{\rho}^{\mathrm{stat}}$,
\emph{i.e.} we start counting the charge passing through the right
junction at $t=0$ and the system is in its stationary state. Now,
combining Eqs. (\ref{eq_resolvent},
\ref{eq_squantity}-\ref{eq_d/dzF}) and having in mind that
$\mathcal{P}\hat{\rho}^{\mathrm{stat}}=\hat{\rho}^{\mathrm{stat}},\mathcal{Q}\hat{\rho}^{\mathrm{stat}}=0$,
straightforward algebra leads to
\begin{equation}
\tilde{S}(\omega) = i\left(\llangle
\tilde{0}|\mathcal{I}_R|0\rrangle-2\llangle
\tilde{0}|\mathcal{I}_R\mathcal{R}(\omega)\mathcal{I}_R|0\rrangle\right).
\end{equation}
We thus arrive at
\begin{equation}
\begin{split}
S_{I_RI_R}(\omega)&=\llangle\tilde{0}|\mathcal{I}_R|0\rrangle-2\mathrm{Re}\left[\llangle\tilde{0}|\mathcal{I}_R\mathcal{R}(\omega)\mathcal{I}_R|0\rrangle\right]\\
&=\llangle\tilde{0}|\mathcal{I}_R|0\rrangle-2\llangle\tilde{0}|\mathcal{I}_R\left[\frac{\mathcal{L}}{\mathcal{L}^2+\omega^2}\right]\mathcal{I}_R|0\rrangle.
\end{split}
\end{equation}
For the left junction one similarly finds
\begin{equation}
S_{I_LI_L}(\omega)=\llangle\tilde{0}|\mathcal{I}_L|0\rrangle-2\mathrm{Re}\left[\llangle\tilde{0}|\mathcal{I}_L\mathcal{R}(\omega)\mathcal{I}_L|0\rrangle\right]
\end{equation}
with
\begin{equation}
\mathcal{I}_L\hat{\rho} =
\Gamma_Le^{-\frac{\hat{x}}{\lambda}}|1\rangle\!\langle
0|\hat{\rho}|0\rangle\!\langle 1|e^{-\frac{\hat{x}}{\lambda}}.
\end{equation}

For the evaluation of the charge-charge correlation function
$S_{NN}(\omega)$ we note that $\hat{N}$ is a system operator, and
the quantum regression theorem thus applies \cite{gardiner_2000}.
Following \cite{flindtPRB_2004} we immediately get
\begin{equation}
S_{NN}(\omega)=-2\mathrm{Re}\left[\llangle\tilde{0}|\mathcal{N}\mathcal{R}(\omega)\mathcal{N}|0\rrangle\right],
\end{equation}
having introduced the superoperator $\mathcal{N}$ corresponding to
$\hat{N}$, defined as
\begin{equation}
\mathcal{N}\hat{\rho}=|1\rangle\!\langle
1|\hat{\rho}|1\rangle\!\langle 1|.
\end{equation}

Collecting all terms in Eq. (\ref{eq_fullnoise}) we finally obtain
the expression for the current noise measured in the leads for the
symmetric setup ($c_L=c_R=\frac{1}{2}$)
\begin{equation}
\begin{split}
S_{II}(\omega)=&\llangle\tilde{0}|
\mathcal{I}_R|0\rrangle+\frac{\omega^2}{2}\mathrm{Re}\left[\llangle\tilde{0}|\mathcal{N}\mathcal{R}(\omega)\mathcal{N}|0\rrangle\right]\\
&-\mathrm{Re}\left[\llangle\tilde{0}|\mathcal{I}_R\mathcal{R}(\omega)\mathcal{I}_R+\mathcal{I}_L\mathcal{R}(\omega)\mathcal{I}_L|0\rrangle\right].
\end{split}
\label{eq_fullnoisefinal}
\end{equation}
We notice that for $\omega=0$ we get the previous result
\cite{novotny_2004,flindtPRB_2004,flindtEPL_2004}
\begin{equation}
S_{II}(0)=\llangle\tilde{0}|\mathcal{I}_R|0\rrangle-2\llangle\tilde{0}|\mathcal{I}_R\mathcal{R}(0)\mathcal{I}_R|0\rrangle,
\end{equation}
since the zero-frequency tunnel current noise is the same at both
junctions and the second term is real \cite{flindtPRB_2004}.

The numerical evaluation of Eq. (\ref{eq_fullnoisefinal}) is only
possible by truncating the number of oscillator states. As in
previous studies \cite{novotny_2003,novotny_2004,flindtEPL_2004}
we retain the 100 lowest oscillator states, which however still
leaves us with the task of dealing numerically with the matrix
representations of the relevant superoperators, which are of size
$20000\times 20000$. As explained in \cite{flindtPRB_2004} the
stationary density matrix $\hat{\rho}^{\mathrm{stat}}$ (or
$|0\rrangle$) can be found using the Arnoldi iteration scheme, and
$\mathcal{R}(\omega)$ can be evaluated using the generalized
minimum residual method (\mbox{GMRes}). Both methods are iterative
and rely crucially on an appropriate choice of preconditioner to
ensure that the iterations converge and to speed up the
computation. It should be stressed that finding a suitable
preconditioner for a given problem is by no means simple. For
finding $\hat{\rho}^{\mathrm{stat}}$ and $\mathcal{R}(0)$ it turns
out that the Sylvester part of the Liouvillean, which is the part
that can be written
$\mathcal{L}_{\mathrm{sylv}}\hat{\rho}=\hat{A}\hat{\rho}+\hat{\rho}\hat{A}^{\dagger}$,
is well-suited for preconditioning \cite{flindtPRB_2004}. This
preconditioner separates the zero eigenvalue from the rest of the
spectrum of $\mathcal{L}$ leading to a decrease in computation
time.

For finding $\mathcal{R}(\omega)$ the original preconditioner must
be modified in order to separate the relevant eigenvalue from the
rest of the spectrum. A reasonable choice is the superoperator
$\mathcal{M}$ defined as
\begin{equation}
\mathcal{M}\hat{\rho}=(\hat{A}+\frac{i\omega}{2})\hat{\rho}+\hat{\rho}(\hat{A}-\frac{i\omega}{2})^{\dagger}.
\end{equation}
For the range of parameters discussed in the present paper this
choice of preconditioner was sufficient for convergence, however
the obtained computational speedup is considerably smaller than
the speedup provided by $\mathcal{L}_{\mathrm{sylv}}$, when
computing $\mathcal{R}(0)$. This observation combined with the
fact that \mbox{GMRes} fails to converge for certain parameters in
the semi-classical regime indicates that the identification of the
optimal preconditioner for the problem at hand remains an open
problem.

\section{Results}

\begin{figure}
  \centering
  \includegraphics[width=0.40\textwidth]{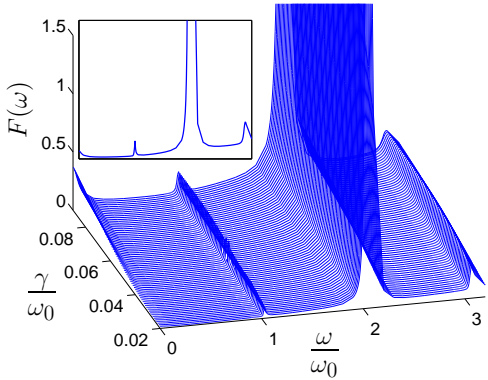}
  \caption{The ratio between the current noise and the current $F(\omega)=S_{II}(\omega)/\langle \hat{I}\rangle$ as function of the damping $\gamma$
  and frequency $\omega$. The other parameters are
  $\Gamma_L=\Gamma_R=0.05\omega_0$, $\lambda=x_0,\, d\equiv eE/m\omega_0^2=0.5x_0$, where $x_0=\sqrt{\hbar/m\omega_0}$.
  Peaks are seen at $\omega\simeq 1.03\omega_0$, $2.06\omega_0, 3.09\omega_0$. The peak at $\omega\simeq 2.06\omega_0$ reaches values of $F(\omega)\simeq 20$ (not shown) for
  $\gamma =0.02$ and decreases monotonously with increasing $\gamma$ to $F(\omega)\simeq 6$ for $\gamma =0.09$. The insert shows a representative curve ($\gamma=0.05$).}
  \label{fig_results1}
\end{figure}

In Fig. \ref{fig_results1} we show results for the current noise
spectrum of the quantum shuttle in the deep quantum regime. In
accordance with previous studies \cite{isacsson_2004,armour_2004}
we find peaks at integer multiples of the mechanical frequency. A
close look at the spectrum reveals a slight renormalization of the
mechanical frequency with the peaks appearing at
$\omega=\tilde{\omega}_0,2\tilde{\omega}_0,3\tilde{\omega}_0$,
where $\tilde{\omega}_0\simeq 1.03\omega_0$. In the shuttling
regime the current is expected to saturate at a value expressed as
one electron per cycle of the mechanical vibrations. For a shuttle
with mechanical frequency $\omega_0$ this implies that the
saturated shuttle current is
$I_{\mathrm{shut}}=\omega_0/2\pi\simeq 0.159\omega_0$. For the
given parameters the numerical calculation yields a slightly
higher value, namely $I_{\mathrm{shut}}=0.164\omega_0=0.159\times
1.03\omega_0\simeq \tilde{\omega}_0/2\pi$, and this can now be
understood in light of the observed renormalization of the
mechanical frequency.

In \cite{flindtEPL_2004,andrea_thesis} it was shown that an
observed giant enhancement of the zero-frequency noise
\cite{novotny_2004} in the coexistence regime of a shuttle
approaching the semi-classical regime can be understood in terms
of a simple model of a bistable system switching slowly between
two current channels (shuttling and tunneling). Denoting the
currents of the two
channels\footnote{$I_{S}=\frac{\tilde{\omega}_0}{2\pi}$, and
$I_{\rm{T}}=\frac{\tilde{\Gamma}_L\tilde{\Gamma}_R}{\tilde{\Gamma}_L+\tilde{\Gamma}_R}$,
with $\tilde{\Gamma}_R=\Gamma_R
e^{\hbar/m\omega_0\lambda^2}e^{2eE/m\omega_0^2\lambda}$,
$\tilde{\Gamma}_L=\Gamma_L e^{\hbar/m\omega_0\lambda^2}$
\cite{novotny_2004,andrea_thesis}.} as $I_{\mathrm{S}}$ and
$I_{\mathrm{T}}$, respectively, and the switching rates
$\Gamma_{S\leftarrow T}$ and $\Gamma_{T\leftarrow S}$, one can
show (following \cite{andrea_thesis}) that the ratio between the
current noise spectrum and the current (in the zero-frequency
limit known as the Fano factor) $F(\omega)=S(\omega)/I$ for the
bistable system has the Lorentzian form
\begin{equation}
F(\omega)=\frac{2}{I}\frac{\Gamma_{S\leftarrow
T}\Gamma_{T\leftarrow S}}{\Gamma_{S\leftarrow
T}+\Gamma_{T\leftarrow S}}\frac{(I_S-I_T)^2}{(\Gamma_{S\leftarrow
T}+\Gamma_{T\leftarrow S})^2+\omega^2}, \label{eq_noisebistable}
\end{equation}
where
\begin{equation}
I=\frac{I_S\Gamma_{S\leftarrow T}+I_T\Gamma_{T\leftarrow
S}}{\Gamma_{S\leftarrow T}+\Gamma_{T\leftarrow S}}.
\end{equation}
The two switching rates, $\Gamma_{S\leftarrow T}$ and
$\Gamma_{T\leftarrow S}$, can be extracted from the numerical
values of current and zero-frequency noise, and by comparing Eq.
(\ref{eq_noisebistable}) with the noise spectrum obtained
numerically, one can perform another independent test of the
hypothesis that the shuttle behaves as a bistable system in the
coexistence regime.

\begin{figure}
  \centering
  \includegraphics[width=0.42\textwidth]{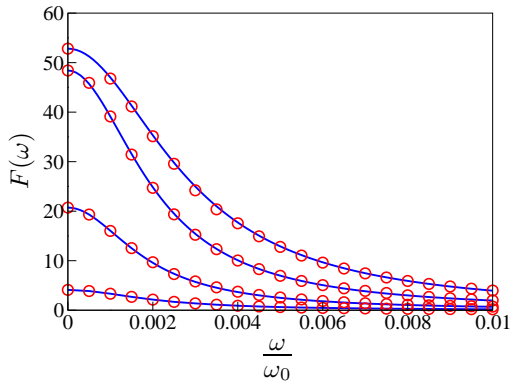}
  \caption{The ratio between the current noise and the current $F(\omega)=S_{II}(\omega)/\langle \hat{I}\rangle$ for low frequencies ($\omega\ll\omega_0$). The parameters
  are $\gamma = 0.035\omega_0$ (lowest curve), $0.04\omega_0,0.045\omega_0, 0.05\omega_0$ (topmost curve), $\Gamma_L=\Gamma_R=0.01\omega_0$, $\lambda=1.5x_0,\, d\equiv eE/m\omega_0^2=0.5x_0$, where $x_0=\sqrt{\hbar/m\omega_0}$. The circles
  indicate numerical results, while the full lines indicate the analytic results for the current noise spectrum of a bistable system. It should be noted that in order to obtain the agreement between the numerical and analytic results it is
  necessary to assume that the shuttling current for the given values of $\gamma$ is not fully saturated to the value $I_{\mathrm{shut}}=1.03\omega_0/2\pi$. Corresponding to the different values of $\gamma$ we have used $I_{\mathrm{shut}}=1.01\omega_0/2\pi (\mathrm{for}\,\, \gamma=0.03\omega_0)$,
   $1.00\omega_0/2\pi$, $0.98\omega_0/2\pi$, $0.94\omega_0/2\pi (\mathrm{for}\,\, \gamma=0.05\omega_0)$, respectively.}
  \label{fig_results2}
\end{figure}

In Fig. \ref{fig_results2} we show numerical results for the
low-frequency current noise of the quantum shuttle in the
coexistence regime. Together with the numerical results we show
the analytic expression for the current noise of the bistable
system (Eq. (\ref{eq_noisebistable})) with rates extracted from
the numerical values of current and zero-frequency noise. It
should be noted that the agreement between the numerical and
analytic results could only be obtained by assuming that the
shuttling current for the given values of the damping is not fully
saturated to the value $I_{\mathrm{shut}}=\tilde{\omega}_0/2\pi$.
The current noise spectrum thus provides us with qualitative
evidence for the shuttle behaving as a bistable system in the
coexistence regime, while it, however,  leaves us with an open
question concerning the saturation of the shuttling current.

\section{Conclusion}

We have presented a theory for the calculation of the current
noise spectrum of a large class of nano-electromechanical systems,
namely those that can be described by a Markovian generalized
master equation. As a specific example we have applied the theory
to a quantum shuttle. For the quantum shuttle numerical
calculations of the current noise spectrum in the deep quantum
regime revealed a slight renormalization of the mechanical
frequency -- this, in turn, explains an observed small deviation
of the shuttle current compared to the expected value given by the
product of the natural mechanical frequency and the electron
charge. When approaching the semi-classical regime the
low-frequency noise served as another evidence for the quantum
shuttle behaving as a bistable system, switching slowly between
two current channels, thus supporting this claim, previously based
on the calculation of the full counting statistics of the quantum
shuttle. The theory presented here has a broad range of
applicability, encompassing the few previous studies of the
current noise spectra of nano-electromechanical systems.

\section{Acknowledgements}

The authors would like to thank A. Donarini, A. Armour, and A.
Isacsson for stimulating discussions.

\end{document}